\documentclass[epj]{svjour}
\usepackage{graphics}
\begin{document}
\title{Probing the high-density behavior of symmetry energy with gravitational waves}
\author{F. J. Fattoyev\inst{1}, W. G. Newton\inst{1} \and Bao-An Li\inst{1}
}                     
%
%
\institute{Department of Physics and Astronomy, Texas A\&M
University-Commerce, Commerce, TX 75429, USA}
\date{Received: date / Revised version: date}
%
\abstract{Gravitational wave (GW) astronomy opens up an entirely new
window on the Universe to probe the equations of state (EOS) of
neutron-rich matter. With the advent of next generation GW
detectors, measuring the gravitational radiation from coalescing
binary neutron star systems, mountains on rotating neutron stars,
and stellar oscillation modes may become possible in the near
future. Using a set of model EOSs satisfying the latest constraints
from terrestrial nuclear experiments, state of the art nuclear
many-body calculations of the pure neutron matter EOS, and
astrophysical observations consistently, we study various GW
signatures of the high-density behavior of the nuclear symmetry
energy, which is considered among the most uncertain properties of
dense neutron-rich nucleonic matter. In particular, we find the
tidal polarizability of neutron stars, potentially measurable in
binary systems just prior to merger, is more sensitive to the high
density component of the nuclear symmetry energy than the symmetry
energy at nuclear saturation density. We also find that the upper
limit on the GW strain amplitude from elliptically deformed stars is
very sensitive to the density dependence of the symmetry energy.
This suggests that future developments in modeling of the neutron
star crust, and direct gravitational wave signals from accreting
binaries will provide a wealth of information on the EOS of
neutron-rich matter. We also review the sensitivity of the $r$-mode
instability window to the density dependence of the symmetry energy.
Whereas models with larger values of the density slope of the
symmetry energy at saturation seem to be disfavored by the current
observational data, within a simple $r$-mode model, we point out
that a subsequent softer behavior of the symmetry energy at high
densities (hinted at by recent observational interpretations) could
rule them in.
} 
\authorrunning{F. J. Fattoyev {\sl et al.}}
\maketitle
\section{Introduction}
\label{intro} Understanding the nature of the neutron-rich nucleonic
matter is one of the fundamental quests of both nuclear physics and
astrophysics \cite{NSACLRP2007}. To fulfill this goal, many
experiments and observations are being carried out or proposed using
a wide variety of advanced new facilities, such as the Facility for
Rare Isotope Beams (FRIB), telescopes operating at a variety of
wavelengths, and more sensitive GW detectors (such as Advanced
LIGO-Virgo). The EOS of neutron-rich matter is
a vital ingredient in the interpretation of the results of these experiments and
observations. Within the parabolic approximation the EOS can be
written in terms of the binding energy per nucleon $E(\rho,\alpha)$
as
\begin{equation}
\label{EOS1} E(\rho,\alpha)=E(\rho, 0) + S(\rho) \alpha^2,
\end{equation}
where $\alpha=(\rho_n-\rho_p)/\rho$ is the isospin asymmetry,
$E(\rho, 0)$ is the binding energy per nucleon in symmetric nuclear
matter (SNM), and $S(\rho)$ is the symmetry energy which represents
the energy cost per nucleon of changing all the protons in SNM into neutrons. Around the saturation density $\rho_0$, one can
make further expansions:
\begin{eqnarray}
&& E(\rho, 0) = B + \frac{1}{2}K_{0} \chi^2 + \mathcal{O}(\chi^3)
\;,
\\ && S(\rho) = J + L \chi + \frac{1}{2}K_{\rm sym}\chi^2 +
\mathcal{O}(\chi^3) \;,
\end{eqnarray}
where $\chi \equiv \left(\rho - \rho_0\right)/3\rho_0$, $B$ and
$K_0$ are the binding energy per nucleon and the nuclear
incompressibility at saturation density, $\rho_0$, while $J$, $L$,
and $K_{\rm sym}$ are the corresponding magnitude, slope, and
curvature of the symmetry energy at saturation density. Whereas
significant progress has been achieved in constraining the EOS of
SNM around $\rho \approx \rho_0$, its density dependence remains
still rather uncertain for nuclear matter at high densities and with
large isospin asymmetries. Observationally, the best constraint on
the high-density component of the EOS comes from the largest
observed masses of neutron stars that have been recently reported by
Demorest {\sl et al.}\,\cite{Demorest:2010bx} and Antoniadis {\sl et
al.}\,\cite{Antoniadis:2013pzd}. Combining observational results
with terrestrial experimental studies of the collective flow and
kaon production in relativistic heavy-ion
collisions\,\cite{Danielewicz:2002pu} the EOS of SNM has been
limited to a relatively small range up to about 4.5 $\rho_0$. The
main source of uncertainties in the EOS of neutron-rich matter
therefore mostly comes from the poorly known density dependence of
the nuclear symmetry energy $S(\rho)$, which is a vital ingredient
in describing the structure of rare isotopes and their reaction
mechanisms. Moreover, it determines uniquely the proton fraction and
therefore the condition for the onset of the direct Urca process in
neutron stars, affects significantly structural properties such as
the radii, moments of inertia and the crust thickness, as well as
the frequencies and damping times of various oscillation modes of
neutron stars (see Refs.
\cite{Lattimer:2000kb,Lattimer:2004pg,Lattimer:2012xj} for review).

Intensive efforts devoted to constraining $S(\rho)$ using various
approaches have recently led to a close
convergence~\cite{Lattimer:2012xj,Xu:2010fh,Newton:2011dw,Steiner:2011ft,Dutra:2012mb,Tsang:2012se,Fattoyev:2012ch}
around $J \approx 30$ MeV and its density slope of $L\approx 60$
MeV, although the associated error-bars from different approaches
may vary broadly. However, the possibility that $J$ and $L$ can be
significantly different from these currently inferred values cannot
be conclusively ruled out (See Ref. \cite{Fattoyev:2013yaa} for a
detailed discussion). On the other hand, the high-density behavior
of $S(\rho)$ remains quite uncertain despite its importance to
understanding what happens in the core of neutron
stars~\cite{Kutschera:1993pe,Kubis:1999sc,Kubis:2002dr,Wen:2009av,Lee:2010sw}
and in reactions with high energy radioactive
beams~\cite{Li:2008gp}. The predictions for the high-density
behavior of the symmetry energy from all varieties of nuclear models
diverge
dramatically~\cite{Brown:2000pd,Szmaglinski:2006fz,Dieperink:2003vs}.
Some models predict very stiff symmetry energies that monotonically
increase with
density~\cite{Dieperink:2003vs,Steiner:2004fi,Lee:1998zzd,Horowitz:2000xj,Chen:2007ih,Li:2006gr},
while others predict relatively soft ones, or an $S(\rho)$ that
first increases with density, then saturates and starts decreasing
with increasing
density~\cite{Brown:2000pd,Pandharipande:1972,Friedman:1981qw,Wiringa:1988tp,Kaiser:2001jx,Krastev:2006ii,Chabanat:1997qh,RikovskaStone:2003bi,Chen:2005ti,Decharge:1979fa,Das:2002fr,Khoa:1996,Basu:2007ye,Myers:1994ek,Banik:1999cb,Chowdhury:2009jn}.
These uncertainties can be associated with our poor knowledge about
the isospin dependence of the strong interaction in the dense
neutron-rich medium, particularly the spin-isospin dependence of
many-body forces, the short-range behavior of the nuclear tensor
force and the isospin dependence of nucleon-nucleon correlations in
the dense medium, see, e.g. Refs.~\cite{Xu:2009bb,Xu:2012hf}. The
experimental progress in constraining the high density $S(\rho)$ is
also limited partially due to the lack of sensitive probes. Whereas
several observables have been proposed~\cite{Li:2008gp} and some
indications of the high-density $S(\rho)$ have been reported
recently~\cite{Xiao:2009zza,Russotto:2011hq}, conclusions based on
terrestrial nuclear experiments remain highly
controversial~\cite{Trautmann:2012nk}. Interestingly, it was
recently proposed that the late time neutrino signal from a core
collapse supernova~\cite{Roberts:2011yw} and the tidal
polarizability~\cite{Fattoyev:2012uu} of canonical neutron stars in
coalescing binaries are very sensitive probes of the high-density
behavior of nuclear symmetry energy, suggesting that astrophysical
measurements might bolster experimental results in this area.

In this survey we focus on the sensitivity of various gravitational
wave signals from neutron stars to the high-density component of the
nuclear symmetry energy. Gravitational waves are oscillations of
space-time and are one of the fundamental predictions of the theory
of general relativity. Although they have not been directly detected
yet, there are strong indirect evidence that gravitational radiation
exists. The highly accurate matching of the observed decrease in the orbital
period of the celebrated Hulse-Taylor~\cite{Hulse:1974eb} and PSR
J0737-3039 binary systems~\cite{Burgay:2003jj,Lyne:2004cj} with the
predicted value due to the energy loss through gravitational
radiation serves as the most precise indirect evidence of the existence
of gravitational waves. Because of their extremely weak interaction
with matter, gravitational waves carry much cleaner information of
their source as opposed to their electromagnetic counterparts, and
therefore open an entirely new window to probe physics that is
hidden to current electromagnetic observations. Of course, this property also makes their detection
one of the most difficult experimental problems faced in physics today.

The manuscript has been organized as follows. In Section 2 we
describe the formalism of constraining the EOS of SNM, and discuss
the emergence of uncertainties in the high-density component of the
EOS of neutron-rich matter due to the density dependence of symmetry
energy. In Section 3 we present results for the selected
gravitational wave signatures from neutron stars that are
investigated consistently with the same set of EOSs discussed in
Section 2. In particular, we study the sensitivity of gravitational
waves generated by ellipticities in the neutron star shape generated
by tidal polarization and mountains of accreted material, and by
$r$-mode oscillations, to the high-density component of the nuclear
symmetry energy. Finally, our concluding remarks are summarized in
Section 4.

\section{Constraining EOS of neutron-rich nuclear matter}
\label{EOSModel} We will concentrate on two models of the nuclear
energy density functionals (EDF): the relativistic mean-field (RMF)
model and the Skyrme Hartree-Fock (SHF)
approach~\cite{Fattoyev:2012ch,Fattoyev:2012uu}. For detailed
discussion on these EDFs see the contributions to this volume by
Nazarewicz {\sl et al.} and by Piekarewicz. All of the EOSs used in
this work are adjusted to satisfy the following four conditions
within their respective known uncertain ranges:
\begin{itemize}
\item[1)] Reproducing the EOS of pure neutron matter (PNM) at
sub-saturation densities predicted by the latest state of the art
microscopic nuclear many-body
calculations~\cite{Friedman:1981qw,Schwenk:2005ka,vanDalen:2009vt,Hebeler:2009iv,Gandolfi:2008id,Gezerlis:2009iw,Vidana:2001ei};
\item[2)] Predicting correctly saturation properties of symmetric nuclear
matter, {\sl i.e.}, nucleon binding energy $B = -16 \pm 1 \, {\rm
MeV}$, incompressibility of nuclear matter $K_{0} = 230 \pm 20 \,
{\rm MeV}$~\cite{Youngblood:1999,Lui:2011zz}, and the nucleon
(Dirac) effective mass $M^{\ast}_{\rm D, 0} = 0.61\pm 0.03 \,
M$~\cite{Furnstahl:1997tk} at saturation density $\rho_0 = 0.155 \pm
0.01 \, {\rm fm}^{-3}$;
\item[3)]  Predicting a fiducial value of the symmetry
energy  $\tilde{J} = 26 \pm 0.5 \, {\rm MeV}$ at a subsaturation
density of $2\rho_0/3$. Notice that theoretical microscopic PNM
calculations also put tight constraints on both the symmetry energy
and its density slope at saturation density: $J =31 \pm 2 \, {\rm
MeV}$ and the density slope of the symmetry energy $L = 50 \pm 10 \,
{\rm MeV}$ (See Ref. ~\cite{Fattoyev:2012ch} and references
therein). Also note that this range is a model dependent prediction
coming from the theoretical constraints of PNM, that has no
experimental analog;
\item[4)] Passing through the terrestrial constraints on the EOS of SNM between $2\rho_0$ and
4.5$\rho_0$~\cite{Danielewicz:2002pu} and predicting a maximum
observed mass of neutron stars of about $2M_{\odot}$ assuming they
are made of the nucleons (neutrons and protons) and light leptons
(electrons and muons) only and without considering other degrees of
freedom or invoking any exotic
mechanism~\cite{Demorest:2010bx,Antoniadis:2013pzd}.
\end{itemize}
\begin{figure}[h]
\vskip -1 cm
\resizebox{0.65\textwidth}{!}{%
\includegraphics{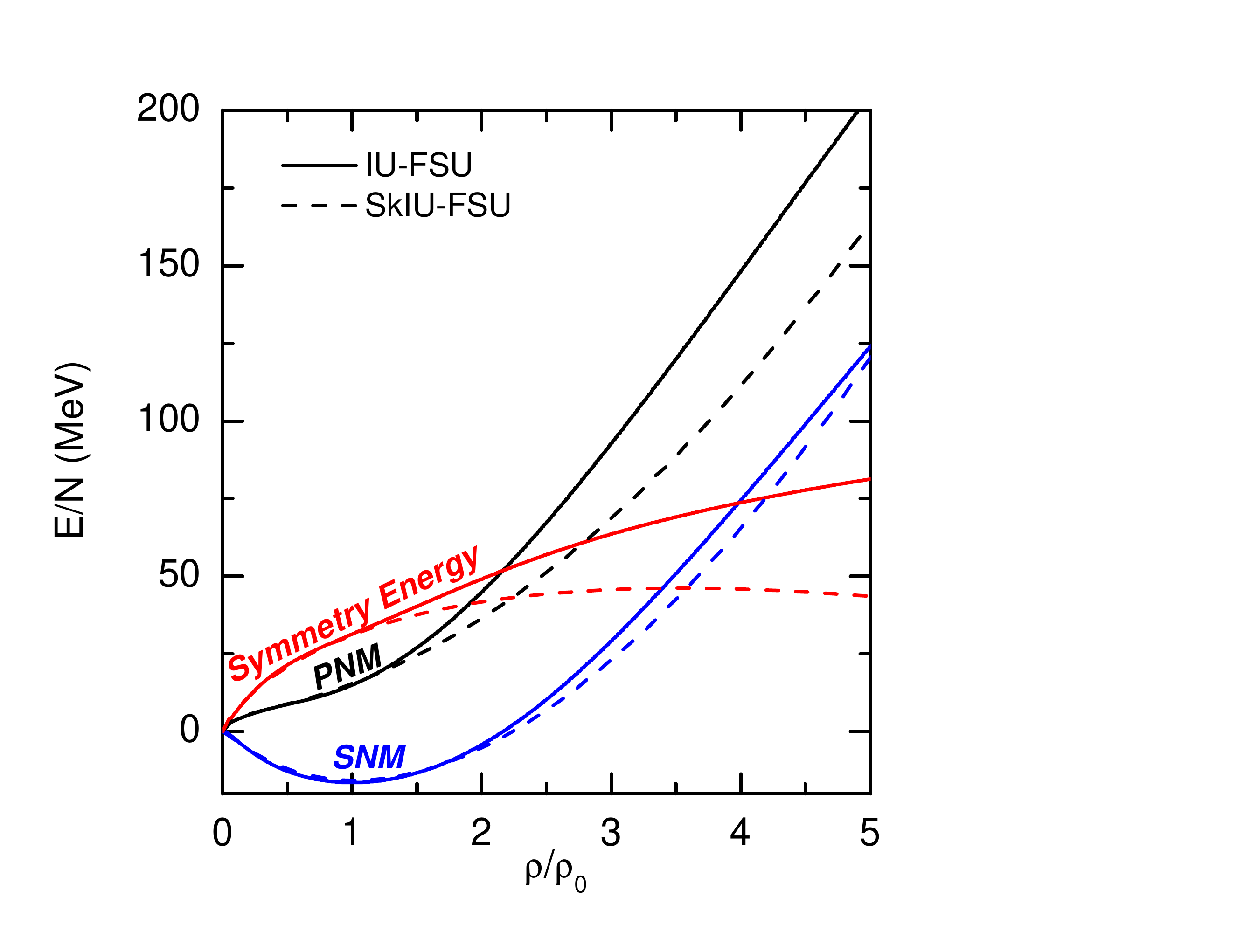}
}
\vskip -0.5 cm \caption{(Color online) The EOS of SNM and PNM as
well as the symmetry energy as a function of density obtained within
the IU-FSU RMF model and the SHF approach using the SkIU-FSU
parameter set. Taken from Ref.~\cite{Fattoyev:2012uu}.}
\label{Fig1}       
\end{figure}

As an example, two EOSs were obtained using the IU-FSU
parametrization of the RMF model~\cite{Fattoyev:2010mx} and its SHF
counterpart dubbed as the SkIU-FSU parameter
set~\cite{Fattoyev:2012ch}, which are shown in Fig.~\ref{Fig1}. By
design, they both have the same EOS for SNM and PNM around and below
$\rho_0$. Thus, at sub-saturation densities the values of $S(\rho)$
which is approximately the difference between the EOSs for PNM and
SNM are almost identical for the two models. However, the values of
$S(\rho)$ are significantly different above about $1.5\rho_0$ with
the IU-FSU leading to a much stiffer $S(\rho)$ at high densities. As
discussed in Ref.~\cite{Fattoyev:2012ch} this is due to the
characteristic differences in the functional forms of the symmetry
energy and is a generic feature of RMF and SHF models:
\begin{eqnarray}
&& \label{SymEnerRMF} S_{\rm RMF}(\rho) = A(\rho) \rho^{2/3} + B(\rho) \rho \ , \\
&& \label{SymEnerSHF} S_{\rm SHF}(\rho) = a \rho^{2/3} - b \rho - c
\rho^{5/3} - d \rho^{\sigma + 1} \ ,
\end{eqnarray}
where $A(\rho)$ and $B(\rho)$ are positive-valued functions of
density, $a$, $b$, $c$, $d$ and $\sigma$ are some constant functions
that may depend on Skyrme parameters only (for example, $\sigma$ is
just a Skyrme parameter). The symmetry energy in the RMF model is
therefore always positive, while certain terms of the symmetry
energy in the SHF model can become negative at higher densities.
More quantitatively, the $S(\rho)$ with IU-FSU is $40$ to $60\%$
higher in the density range of $3 \rho_0$ to $4\rho_0$ expected to
be attained in the core of canonical neutron stars.
\begin{figure}[h]
\resizebox{0.5\textwidth}{!}{%
\includegraphics{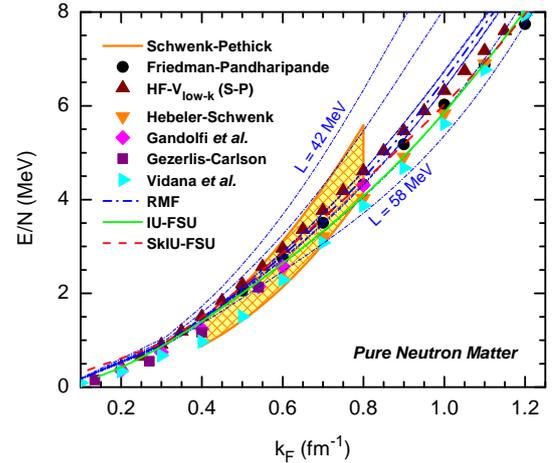}
}
\caption{(Color online) Energy per nucleon as a function of the
Fermi momentum for PNM for selected models described in the text
(Taken from Ref.~\cite{Fattoyev:2012uu}).} \label{Fig2}
\end{figure}
\begin{figure}[h]
\resizebox{0.5\textwidth}{!}{%
\includegraphics{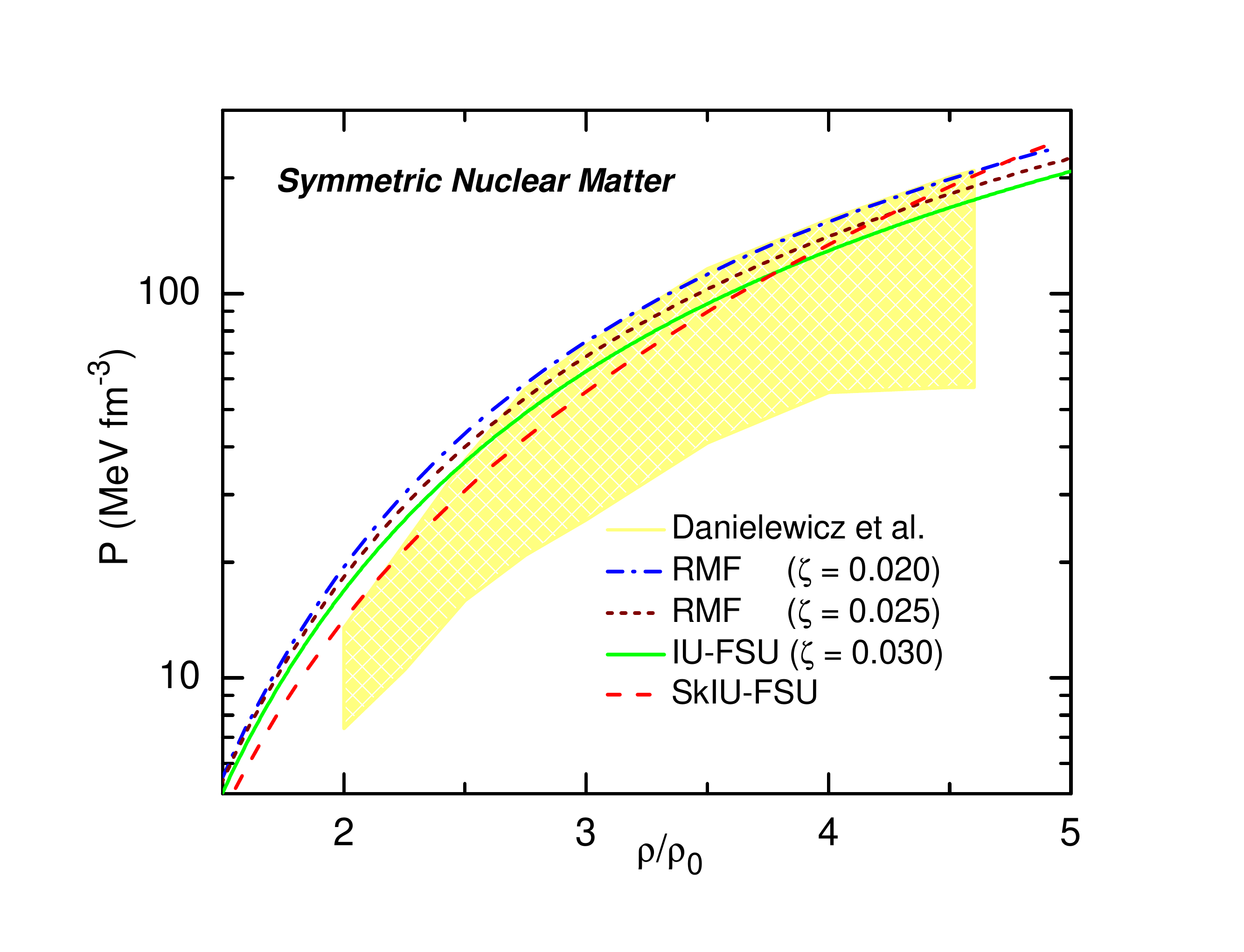}
}
\caption{(Color online) The pressure of SNM as a function of
baryon density.  Here $\rho_0$ is the nuclear matter saturation
density and the shaded area represents the EOS extracted from the
analysis of \cite{Danielewicz:2002pu}. The figure is taken from
Ref.~\cite{Fattoyev:2012uu}.} \label{Fig3}
\end{figure}

To test the sensitivity of the various gravitational wave
observables to variations of properties of neutron-rich nuclear
matter around $\rho_0$ within the constraints listed above, we also
build seventeen different RMF parameterizations by systematically
varying the values of $K_0$, $M_{\rm 0}^{\ast}$, $L$, and the
$\zeta$ parameter of the RMF model that controls the quartic
omega-meson self interactions~\cite{Mueller:1996pm} and subsequently
the high-density component of the EOS of SNM (See Table \ref{Tab1}
for their predictions). Besides the constraints listed above, all
parametrizations can correctly reproduce the experimental values for
the binding energy and charge radius of $^{208}$Pb and the ground
state properties of other doubly magic nuclei within 2\%
accuracy~\cite{Todd:2003xs}. As a reference for comparisons, we
select $K_0 = 230$ MeV, $M_0^{\ast} = 0.61$ $M$, $L =50$ MeV, and
$\zeta = 0.025$ which we refer to as our baseline model. This model
predicts $\rho_0 = 0.1524$ fm$^{-3}$, $B = -16.33$ MeV and $J =
31.64$ MeV. The representative model EOSs for PNM at sub-saturation
densities and those for SNM at supra-saturation densities are then
compared with their constraints in Fig.~\ref{Fig2} and
Fig.~\ref{Fig3}, respectively. It is seen that the SkIU-FSU and all
the RMF models with $42< L< 58$ MeV can satisfy the constraint from
the PNM EOS. Also, they all simultaneously satisfy the high density
SNM EOS constraint with $0.02< \zeta< 0.03$. Moreover, they all give
a maximum mass for neutron stars in a range between $1.94 M_{\odot}$
and $2.07 M_{\odot}$~\cite{Fattoyev:2010mx}, the upper end of which
is consistent with existing precise observational
measurements~\cite{Demorest:2010bx,Antoniadis:2013pzd}. While the
determination of neutron-star radii from observations is both
challenging and hindered by uncertainties in models of the stellar
atmosphere amongst other things, significant advances in X-ray
astronomy have allowed the simultaneous determination of masses and
radii from a systematic study of several X-ray bursters. By assuming
that all neutron stars have the same radius Guillot {et.
al}~\cite{Guillot:2013} have recently analyzed the thermal spectra
of 5 quiescent low-mass X-ray binaries in globular clusters to find
that neutron star have radii of $R=9.1^{+1.3}_{-1.5}$ km at a 90\%
confidence level. A subsequent re-analysis using Bayesian approach
by Lattimer and Steiner~\cite{Lattimer:2013} suggests that different
interpretations of the data is strongly favored and they find much
larger neutron star radii of $R=12.1^{+0.7}_{-0.7}$ km for a 1.4
solar mass neutron star; neither range can currently be conclusively
ruled out. The models discussed in our text predict canonical
neutron star radii between 12.33 km and 13.22 km in the range of
currently observed
values~\cite{Guillot:2013,Lattimer:2013,Steiner:2010fz,Ozel:2010fw,Suleimanov:2010th}.

\section{Gravitational waves signals}
\label{GWs}
\subsection{Tidal Love number and polarizability}
\label{tidal} Coalescing binary neutron stars are among the most
promising sources of laser-interferometric gravitational waves. One
of the most important features of binary mergers is the tidal
deformation neutron stars experience as they approach each other
prior to merger. The strength of the tidal deformation can give us
invaluable information about the neutron-star matter EOS
\cite{Flanagan:2007ix,Hinderer:2007mb,Binnington:2009bb,Damour:2009vw,Damour:2009wj,Hinderer:2009ca,Postnikov:2010yn,Baiotti:2010xh,Baiotti:2011am,Lackey:2011vz,Pannarale:2011pk,Damour:2012yf}.
At the early stage of an inspiral tidal effects may be effectively
described through the tidal polarizability parameter $\lambda$
\cite{Flanagan:2007ix,Hinderer:2009ca,Damour:2009vw,Damour:2009wj}
defined via
\begin{equation}
Q_{ij}= - \lambda \mathcal{E}_{ij} ,
\end{equation}
where $Q_{ij}$ is the induced quadrupole moment of a star in binary
due to the static external tidal field of the companion star
$\mathcal{E}_{ij}$. The tidal polarizability can be expressed in
terms of the dimensionless quadrupolar tidal Love number $k_2$:
\begin{equation}
\lambda = \frac{2}{3}R^5k_2 \ ,
\end{equation}
where $R$ is the radius of a neutron star in isolation, {\sl i.e.}
long before the merger. The tidal Love number $k_2$ depends on the
stellar structure and can be calculated using the following
expression~\cite{Hinderer:2007mb,Postnikov:2010yn}:
\begin{eqnarray} \nonumber
k_2 &=& \frac{1}{20}\left(\frac{R_s}{R}\right)^5 \left(1-
\frac{R_s}{R}\right)^2\left[2- y_{R} + \left(y_R-1\right)
\frac{R_s}{R} \right] \times \\
\nonumber &\times& \bigg\{\frac{R_s}{R} \bigg(6 - 3 y_R +
\frac{3R_s}{2R} \left(5y_R-8\right) +
\frac{1}{4}\left(\frac{R_s}{R}\right)^2 \bigg[26 -\\
\nonumber &-& 22y_R + \left(\frac{R_s}{R}\right) \left(3y_R-2\right)
+ \left(\frac{R_s}{R}\right)^2 \left(1+ y_R\right) \bigg]\bigg) +\\
\nonumber &+& 3\left(1-\frac{R_s}{R}\right)^2\left[2- y_{R} +
\left(y_R-1\right) \frac{R_s}{R} \right] \times \\
 &\times& \log \left(1- \frac{R_s}{R}\right)\bigg\}^{-1} \label{TidalLove} \ ,
\end{eqnarray}
where $R_s \equiv 2 M$ is the Schwarzschild radius of the star, and
$y_R \equiv y(R)$ can be calculated by solving the following
first-order differential equation:
\begin{eqnarray}
r \frac{d y(r)}{dr} + {y(r)}^2 + y(r) F(r) + r^2 Q(r) = 0
\label{TidalLove2} \ ,
\end{eqnarray}
with
\begin{eqnarray}
F(r) = \frac{r-4 \pi r^3 \left( \mathcal{E}(r) - P(r)\right) }{r-2
M(r)} \ ,
\end{eqnarray}
\begin{eqnarray}
\nonumber Q(r) &=& \frac{4 \pi r \left(5 \mathcal{E}(r) +9 P(r) +
\frac{\mathcal{E}(r) + P(r)}{\partial P(r)/\partial
\mathcal{E}(r)} - \frac{6}{4 \pi r^2}\right)}{r-2M(r)} - \\
&-&  4\left[\frac{M(r) + 4 \pi r^3
P(r)}{r^2\left(1-2M(r)/r\right)}\right]^2 \ .
\end{eqnarray}
The Eq. (\ref{TidalLove2}) must be integrated together with the
Tolman-Oppenheimer-Volkoff (TOV) equation. That is,
\begin{eqnarray}
  && \frac{dP(r)}{dr} = - \frac{\Big({\mathcal E}(r)+P(r)\Big)
      \Big(M(r)+4\pi r^{3}P(r)\Big)}{r^{2}\Big(1-2M(r)/r\Big)} \;, \\
  && \frac{dM(r)}{dr} = 4\pi r^{2} {\mathcal E}(r) \;.
 \label{TOV}
\end{eqnarray}
Given the boundary conditions in terms of $y(0) = 2$,
$P(0)\!=\!P_{c}$ and $M(0)\!=\!0$, the tidal Love number can be
obtained once an EOS is supplied. Previous studies have used both
polytropic EOSs and several popular nuclear EOSs available in the
literature~\cite{Flanagan:2007ix,Hinderer:2007mb,Binnington:2009bb,Damour:2009vw,Damour:2009wj,Hinderer:2009ca,Postnikov:2010yn,Baiotti:2010xh,Baiotti:2011am,Lackey:2011vz,Pannarale:2011pk,Damour:2012yf}.
While other particles may be present, it is sufficient to assume
that neutron stars consist of only neutrons (n), protons (p),
electrons (e) and muons $(\mu)$ in
$\beta$-equilibrium~\cite{Fattoyev:2012uu}.
\begin{figure}[h]
\resizebox{0.5\textwidth}{!}{%
\includegraphics{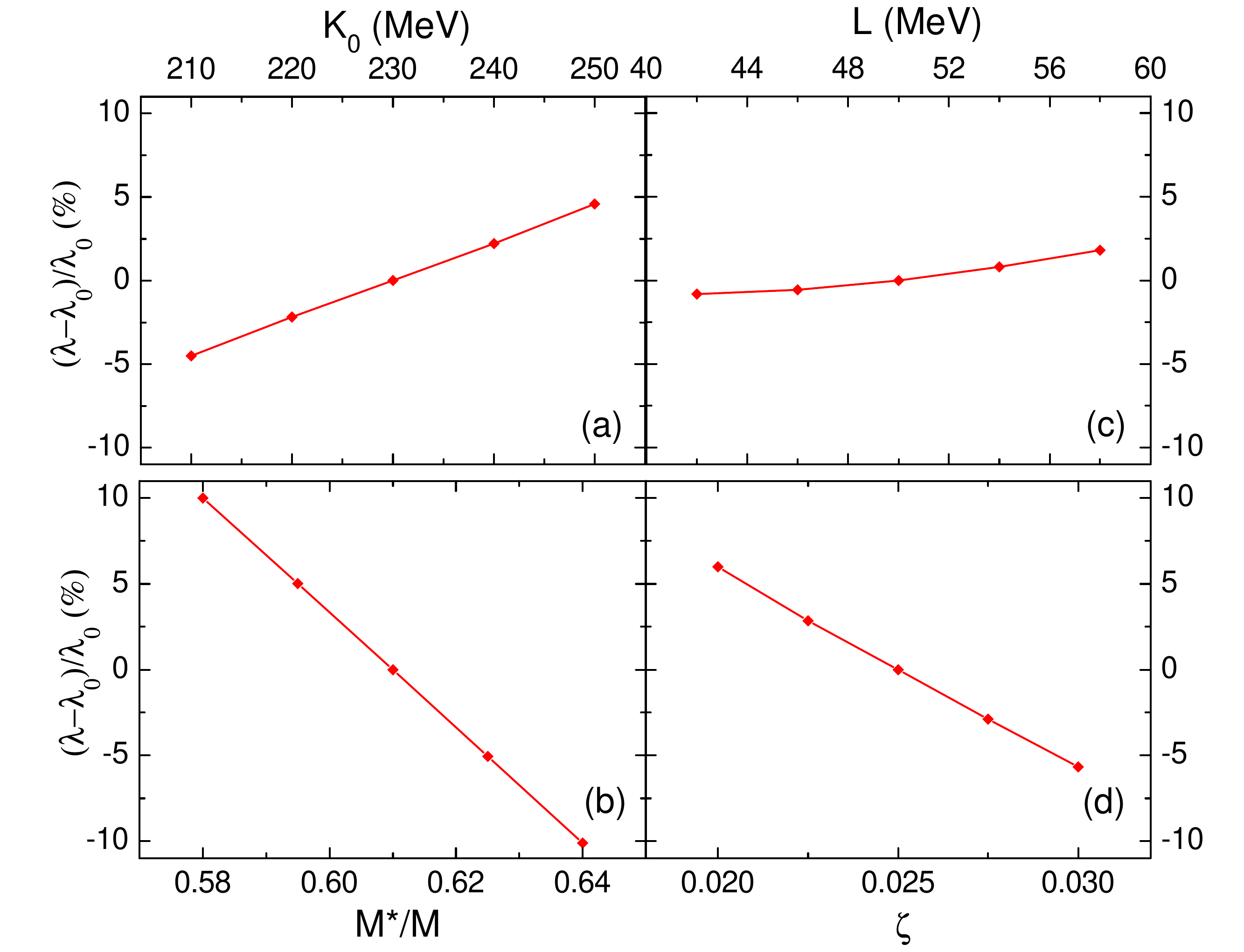}
}
\caption{(Color online) Percentage changes in the tidal
polarizability of a canonical 1.4 solar mass neutron star by
individually varying properties of nuclear matter $K_0$ (a),
$M^{\ast}$ (b), $L$ (c), and the $\zeta$ parameter (d) of the RMF
model with respect to the value using the base model. The figure is
taken from Ref.~\cite{Fattoyev:2012uu}.} \label{Fig4}
\end{figure}

\begin{figure}[h]
\resizebox{0.5\textwidth}{!}{%
\includegraphics{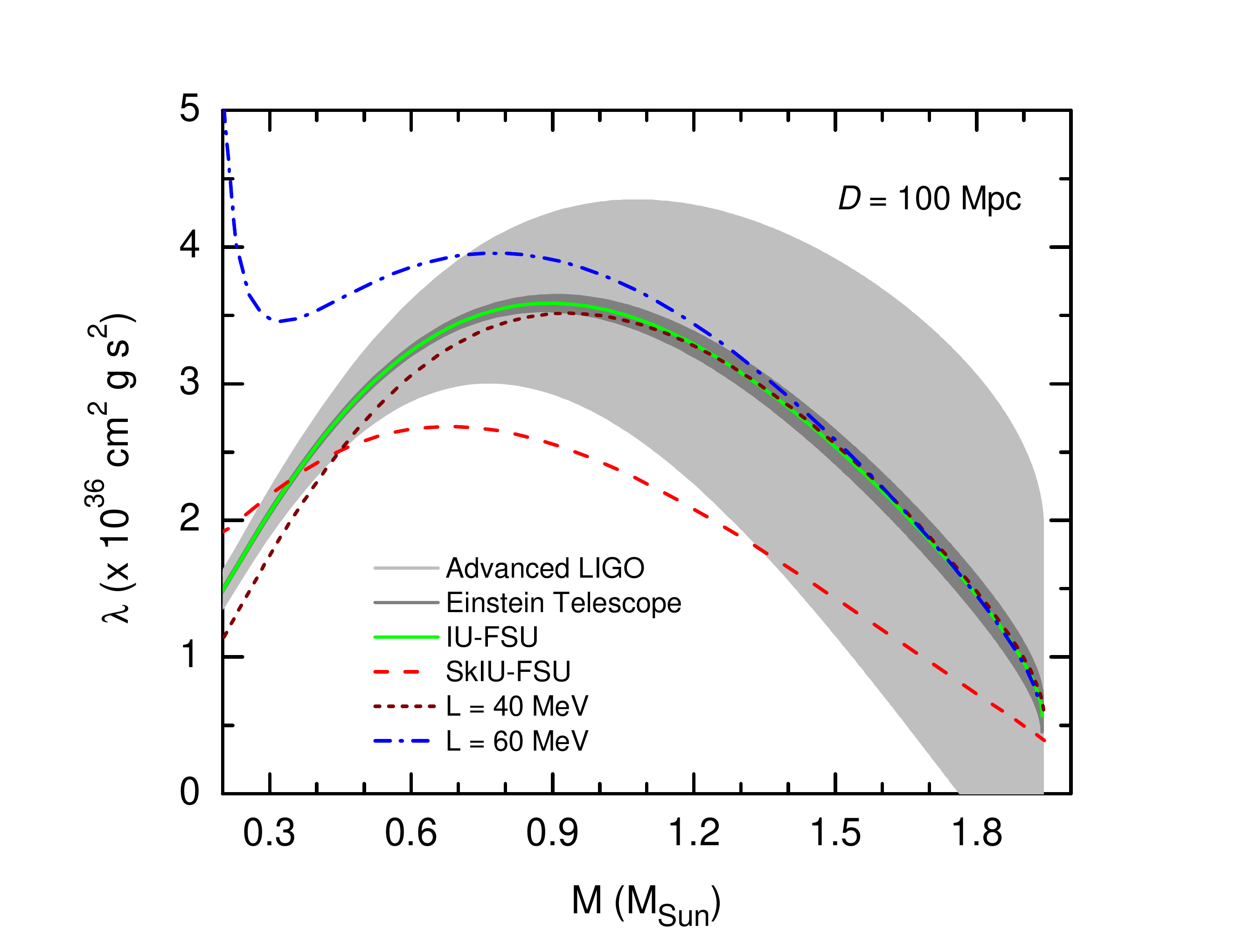}
}
\caption{(Color online) Tidal polarizability $\lambda$ of a single
neutron star as a function of neutron-star mass for a range of EOS
that allow various stiffness of symmetry energies. A crude estimate
of uncertainties in measuring $\lambda$ for equal mass binaries at a
distance of $D = 100$ Mpc is shown for the Advanced
LIGO~\cite{AdvLigo} (shaded light-grey area) and the Einstein
Telescope~\cite{EinsteinTelescope} (shaded dark-grey area). This
result was first reported in Ref.~\cite{Fattoyev:2012uu}.}
\label{Fig5}
\end{figure}

\begin{table}
\begin{center}
\caption{Predictions for the properties of a 1.4 solar mass neutron
star using the seventeen EOSs considered in this paper.The
properties of nuclear matter around our base parametrization are
systematically varied to obtain flexible range of the EOSs, but
within the available theoretical, experimental and observational
constraints. The first column reports the name of the EOS with a
particular nuclear property and/or $\zeta$-parameter indicated. The
radii are in units of km, the tidal polarizability in $10^{36}$
cm$^2$ g s$^{2}$.} \label{Tab1}
\begin{tabular}{@{}lcccccc@{}}
\hline \hline
EOS & $R$  & $k_2$ & $\lambda$ & $\Delta \lambda/\lambda $    \\
\hline \hline
Base          & 12.88  & 0.0879 & 3.115 & ---         \\
\hline
$K = 210$ MeV & 12.82  & 0.0858 & 2.974 & $-$4.52 \%  \\
$K = 220$ MeV & 12.85  & 0.0869 & 3.046 & $-$2.19 \%  \\
$K = 240$ MeV & 12.91  & 0.0890 & 3.183 & $+$2.21 \%  \\
$K = 250$ MeV & 12.94  & 0.0900 & 3.258 & $+$4.59 \%  \\
\hline
$M^{\ast} = 0.580$ $M$ & 12.71  & 0.1033 & 3.427 & $+$10.01 \%  \\
$M^{\ast} = 0.595$ $M$ & 12.83  & 0.0943 & 3.271 & $+$~5.02 \%  \\
$M^{\ast} = 0.625$ $M$ & 12.89  & 0.0831 & 2.957 & $-$~5.06 \%  \\
$M^{\ast} = 0.640$ $M$ & 12.88 & 0.0792 & 2.800 & $-$10.12 \%  \\
\hline
$L = 42$ MeV & 12.33 & 0.1677  & 3.089 & $-$0.83 \%  \\
$L = 46$ MeV & 12.64 & 0.1635 & 3.096 & $-$0.58 \% \\
$L = 54$ MeV & 13.07 & 0.1582 & 3.140 & $+$0.80 \%  \\
$L = 58$ MeV & 13.22 & 0.1564  & 3.170 & $+$1.79 \%  \\
\hline
$\zeta = 0.0200$ & 13.01 & 0.0885 & 3.302 & $+$6.00 \%  \\
$\zeta = 0.0225$ & 12.94  & 0.0882 & 3.204 & $+$2.85 \%  \\
$\zeta = 0.0275$ & 12.81 & 0.0876 & 3.025 & $-$2.90 \%  \\
$\zeta = 0.0300$ & 12.75 & 0.0873 & 2.938 & $-$5.67 \%  \\
\hline \hline
\end{tabular}
\end{center}
\end{table}

\begin{table}
\begin{center}
\caption{Predictions for the properties of a 1.4 solar mass neutron
star using the IU-FSU EOS with difference density dependence of the
symmetry energy. The slopes of the symmetry energy are in units of
MeV, radii are in units of km, and the tidal polarizability in
$10^{36}$ cm$^2$ g s$^{2}$. The relative percentage error $\Delta
\lambda/\lambda $ is calculated with respect to the original IU-FSU
parametrization~\cite{Fattoyev:2010mx}.} \label{Tab2}
\begin{tabular}{@{}lccccccc@{}}
\hline \hline
EOS     & $L$ & $R$  & $k_2$ & $\lambda$ & $\Delta \lambda/\lambda $    \\
\hline \hline
IU-FSU      & 47.2   & 12.49 & 0.0930 & 2.828 & ---         \\
IU-FSU-min  & 40.0   & 12.20  & 0.1054 & 2.841 & $+$~0.46 \%  \\
IU-FSU-max  & 60.0   & 13.07  & 0.0761 & 2.906 & $+$~2.76 \%  \\
SkIU-FSU    & 47.2 & 11.71  & 0.0753 & 1.657 & $-$41.41 \%  \\
\hline \hline
\end{tabular}
\end{center}
\end{table}

In Ref.~\cite{Fattoyev:2012uu}, we examined sensitivity of the tidal
polarizability $\lambda$ of a $1.4M_{\odot}$ neutron star to the
variations of SNM properties and the slope of the symmetry energy
around $\rho_0$ as shown in Fig.~\ref{Fig4} and Table~\ref{Tab1}.
The changes of $\lambda$ relative to the values for our base RMF
model are shown for the remaining RMF EOSs. It is very interesting
to see that the tidal polarizability is rather insensitive to the
variation of $L$ within the constrained range, although it changes
up to $\pm 10\%$ with $K_0$, $M^{\ast}$ and $\zeta$ within their
individual uncertain ranges.

While the averaged mass is $M = 1.33
\pm 0.05 \, M_{\odot}$, neutron stars in binaries have a broad mass
distribution~\cite{Ozel:2012ax}. It is thus necessary to investigate
the mass dependence of the tidal polarizability. Whereas what can be
measured for a neutron star binary of mass $M_1$ and $ M_2$ is the
mass-weighted tidal polarizability~\cite{Hinderer:2009ca}
\begin{equation}
\tilde{\lambda} = \frac{1}{26} \left[\frac{M_1+12 M_2}{M_1}
\lambda_1  + \frac{M_2+12 M_1}{M_2} \lambda_2  \right] \ ,
\end{equation}
for the purpose of this study it is sufficient to consider binaries
consisting of two neutron stars with equal masses. In
Fig.~\ref{Fig5} the tidal polarizability $\lambda$ as a function of
neutron-star mass for a range of EOSs is shown. Very interestingly,
it is seen that the IU-FSU and SkIU-FSU models which are different
only in their predictions for the nuclear symmetry energy above
about $1.5\rho_0$ (see Fig.~\ref{Fig1}) lead to significantly
different $\lambda$ values in a broad mass range from 0.5 to 2
$M_{\odot}$. More quantitatively, a $41\%$ change in $\lambda$ from
$2.828\times 10^{36}$ (IU-FSU) to $1.657\times 10^{36}$ (SkIU-FSU)
is observed for a canonical neutron star of 1.4 $M_{\odot}$ (See
Table~\ref{Tab2}). For a comparison, we notice that this effect is
as strong as the symmetry energy effect on the late time neutrino
flux from the cooling of proto-neutron stars~\cite{Roberts:2011yw}.
However, we should note that the SNM components of the EOSs used in
Ref.~\cite{Roberts:2011yw} are also significantly different and
therefore a further systematic test may be desirable. Moreover, it
is shown that the variation of $L$ on 40 to 60 MeV as allowed by the PNM constraints has a very small effect on the
tidal polarizability $\lambda$ of massive neutron stars, which is
consistent with the results shown in Fig.~\ref{Fig4}. On the other
hand, the $L$ parameter affects significantly the tidal
polarizability of neutron stars with $M\leq 1.2M_{\odot}$. These
observations can be easily understood. From Eq. (\ref{TidalLove})
the Love number $k_2$ is essentially determined by the compactness
parameter $M/R$ and the function $y(R)$. Both of them are obtained
by integrating the EOS all the way from the core to the surface.
Since the saturation density approximately corresponds to the
central density of a $0.3M_{\odot}$ neutron star, one thus should
expect that only the Love number of low-mass neutron stars to be
sensitive to the EOS around the saturation density. However, for
canonical and more massive neutron stars, the central density is
higher than $3$ to $4$ saturation density $\rho_0$, and therefore
both the compactness $M/R$ and $y(R)$ show stronger sensitivity to
the variation of EOS at supra-saturation densities. Since all the
EOSs for SNM at supra-saturation densities have already been
constrained by the terrestrial nuclear physics data and required to
give a maximum mass about $2M_{\odot}$ for neutron stars, the
strongest effect on calculating the tidal polarizability of
massive neutron stars should therefore come from the high-density
behavior of the symmetry energy.

It has been suggested that the Advanced LIGO-Virgo detector may
potentially measure the tidal polarizability of binary neutron stars
with a moderate accuracy. To test whether planned GW detectors are
sensitive enough to measure the predicted effects of high-density
symmetry energy on the tidal polarizability, as an example, we
estimate uncertainties in measuring $\lambda$ for equal mass
binaries at an optimally-oriented distance of $D = 100$
Mpc~\cite{Hinderer:2009ca,Abadie:2010cf} using the same approach as
detailed in Refs.~\cite{Hinderer:2009ca,Damour:2012yf}. For example,
for point-particle models of binary inspiral,
Ref.~\cite{Flanagan:2007ix} showed that one can obtain analytical
gravitational waveform accurate to 2-3 post-Newtonian (PN) order.
The tidal contribution to the GW signal is then found to be accurate
to less than 10 \%. Current uncertainties in the determination of
$\lambda$ is estimated as~\cite{Hinderer:2009ca}:
\begin{equation}
\label{Hinderer} \Delta \tilde{\lambda} = \alpha
\left(\frac{M}{M_{\odot}}\right)^{2.5}
\left(\frac{M_2}{M_1}\right)^{0.1} \left(\frac{f_{\rm end}}{\rm
Hz}\right)^{-2.2} \left(\frac{D}{\rm 100 \, Mpc}\right) ,
\end{equation}
where $M = M_1 + M_2$ is the total mass of the binary, $\alpha = 1.0
\times 10^{42}$ cm$^2$ g s$^{2}$ for a single Advanced LIGO-Virgo
detector, and $\alpha = 8.4 \times 10^{40}$ cm$^2$ g s$^{2}$ for a
single Einstein Telescope detector. These uncertainties are shown
for the Advanced LIGO-Virgo (shaded light-grey area) and the
Einstein Telescope (shaded dark-grey area) in Fig.~\ref{Fig5}
assuming the IU-FSU as the base model.

It is seen that discerning between high-density symmetry energy
behaviors is at the limit of Advanced LIGO-Virgo's sensitivity for
stars of mass $1.4 M_{\odot}$ and below based on the currently
estimated uncertainty, and it is possible that a rare but nearby
binary system may be found and provide a much more tighter
constraint~\cite{Hinderer:2009ca}. While discoveries of binary
neutron star systems PSR B1913+16 and PSR J0737-3039 at a nearby
location (6.4 kpc and 0.6 kpc respectively) reminds us that such a
possibility may not be likely, the rate estimates for detection of
binary neutron stars are found to be very small for a single
Advanced LIGO-Virgo interferometer~\cite{Hinderer:2009ca}.
Nonetheless, measurements for binaries consisting of light neutron
stars can still help further constrain the symmetry energy around
the saturation density. On the other hand, it is noteworthy that the
narrow uncertain range for the proposed Einstein Telescope even at
very large distances will enable it to stringently constrain the
symmetry energy especially at high densities.

\subsection{Gravitational waves from elliptically deformed pulsars}
\label{elliptical} Rotating neutron stars are major candidates for
sources of continuous gravitational waves in the frequency bandwith
of the current laser interferometric detectors~\cite{Abadie:2010cf}.
According to general relativity any rotating axial-asymmetric
objects should radiate gravitationally. There are several mechanisms
that may lead to an axial asymmetry in neutron
stars~\cite{Krastev:2008yc}:
\begin{itemize}
\item[a)]  Anisotropic stress built up during the crystallization
period of the solid neutron star crust may support static
``mountains" on the surface of neutron stars~\cite{Horowitz:2009ya};
\item[b)]   Due to its violent formation in supernova the rotational axis
of a neutron star may not necessarily coincide with the principal
axis of the moment of inertia, which results in a neutron star
precession~\cite{Zimmermann:1979,Zimmermann:1980};
\item[c)]  Since
neutron stars possess strong magnetic fields they can create a
magnetic pressure, which in turn may distort the star. This is
possible only if the magnetic field axis is not aligned with the
axis of rotation, which is always the case for
pulsars~\cite{Bonazzola:1996}.
\end{itemize}

These processes generally result in a triaxial neutron star
configuration. Gravitational waves are then characterized by a tiny
dimensionless strain amplitude, $h_0$, which depends on the degree
to which the neutron star is deformed from axial
symmetry~\cite{Haskell:2007}:
\begin{equation}
h_0 = \frac{16 \pi^2 G}{c^4} \frac{\epsilon I_{zz} f^2}{r} \ ,
\end{equation}
where $f$ is the rotation frequency of the neutron star, $I_{zz}$ is
its moment of inertia around the principal axis, $\epsilon = (I_{xx}
- I_{yy})/I_{zz}$ is its equatorial ellipticity as defined in the literature on
gravitational waves~\cite{Abadie:2010cf}, and $r$ is the distance
from the source to the observer. The ellipticity is related to the
neutron star maximum quadrupole moment through~\cite{Owen:2005}
\begin{equation}
\epsilon = \sqrt{\frac{8 \pi}{15}} \frac{Q_{22, \rm max}}{I_{zz}} \
.
\end{equation}
Notice that the gravitational wave strain amplitude does not depend
on the neutron star moment of inertia $I_{zz}$, and the total
dependence upon the underlying EOS is entirely due to the maximum
quadrupole moment, $Q_{22, \rm max}$. Using a chemically detailed
model of the crust, Ref.~\cite{Ushomirsky:2000} computes the maximum
quadrupole moment for a neutron star
\begin{equation}
\label{quadrupole} Q_{22, \rm max} = 2.4 \times 10^{38} {\rm
g}\,{\rm cm}^2 \left(\frac{\sigma_{\rm
max}}{10^{-2}}\right)\left(\frac{R}{10\,{\rm km}}\right)^{6.26}
\left(\frac{1.4\,M_{\odot}}{M}\right)^{1.2} \ ,
\end{equation}
where $\sigma_{\rm max}$ is the breaking strain of the crust.
Although earlier studies have estimated the value of the breaking
strain to be in the range of $\sigma_{\rm max}= \left[10^{-5} -
10^{-2}\right]$~\cite{Haskell:2007}, more recently using molecular
dynamics simulations it was estimated that the breaking strain can
be as large as $\sigma_{\rm max} \approx 0.1$, which is considerably
larger than the previous findings \cite{Horowitz:2009ya}. Using a
rather conservative value of $\sigma_{\rm max} = 0.01$, in a work
involving one of us~\cite{Krastev:2008yc} we reported the first
direct nuclear constraints on the gravitational wave signals to be
expected from several pulsars. Particularly, it has been found that
for several millisecond pulsars located at distances $0.18$ kpc to
$0.35$ kpc from Earth, the maximal gravitational wave strain
amplitude is in the range of $h_0 \sim [0.4 - 1.5] \times 10^{-24}$.
The EOS used in Ref.~\cite{Krastev:2008yc} was calculated using the
momentum dependent interaction (MDI). The high density component of
the symmetry energy in MDI is controlled by a single parameter $x$
that is introduced in the single-particle potential of the MDI EOS.
In calculating boundaries of the possible neutron star
configurations, only EOSs that are consistent with the isospin
diffusion laboratory data and measurements of the neutron skin
thickness in $^{208}$Pb~\cite{Li:2005jy,Li:2005sr} were used that
suggest the $x$-parameter is in the range of $x=-1$ (stiff) and
$x=0$ (soft). It was then shown that EOSs with stiff symmetry
energy, such as MDI with $x=-1$, results in less compact stellar
models, and hence more deformed neutron stars. It is of course
reasonable to expect that more compact stellar configurations as
predicted by the models with softer symmetry energy would be more
resistant to any kind of deformation~\cite{Krastev:2008yc}. Also, in
a work involved two of us~\cite{Gearheart:2011} we have demonstrated
that pasta phases can play essential role on the maximum quadrupole
ellipticity sustainable by the crust. In particular, for EOSs with
density slope of the symmetry energy of $L < 70$ MeV, depending on
the pasta phases the effect on the maximum quadrupole ellipticity
can be as large as an order of magnitude.

\begin{figure}[h]
\resizebox{0.5\textwidth}{!}{%
\includegraphics{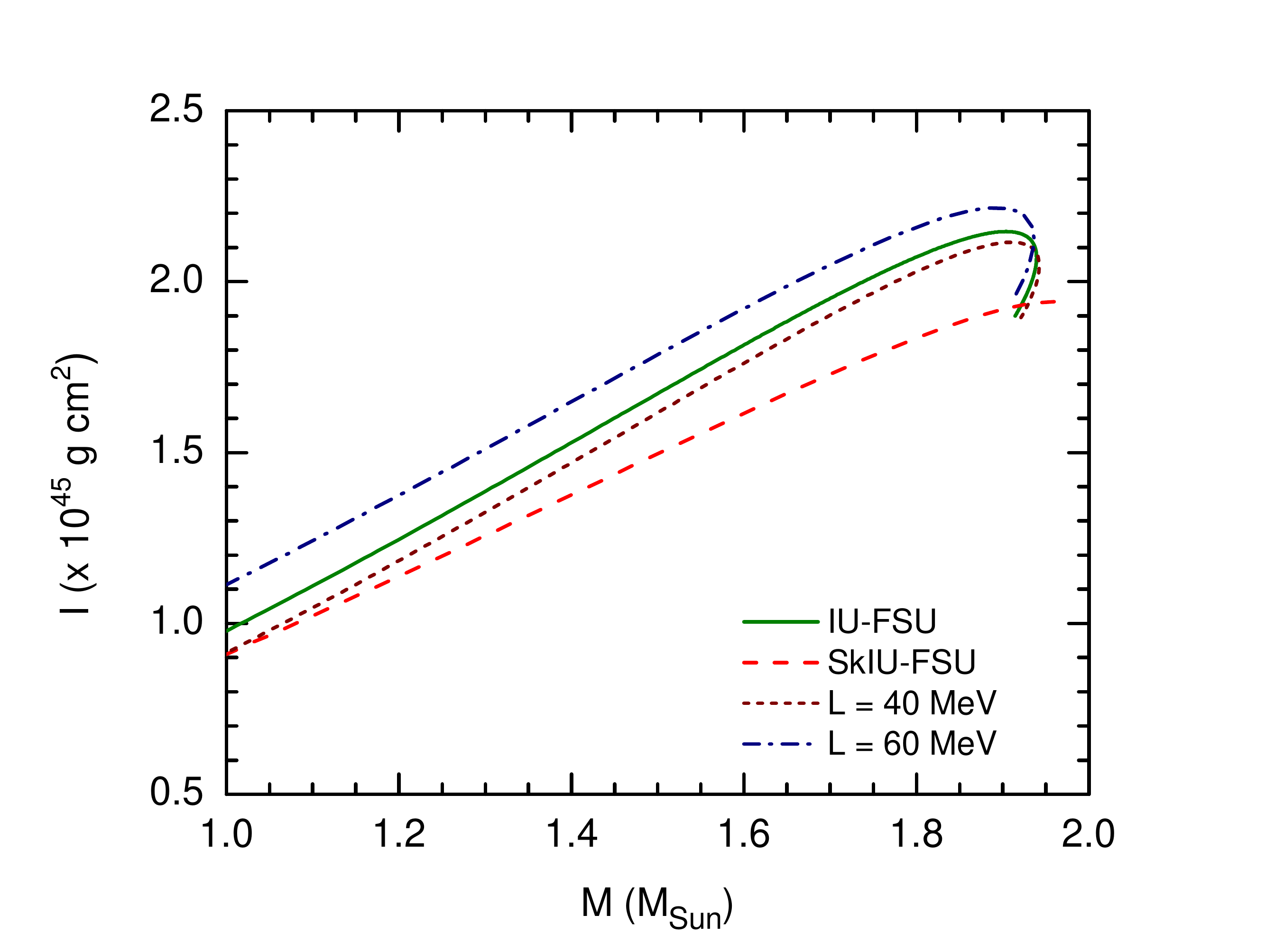}
}
\caption{(Color online) Neutron star moment of inertia as calculated
using Eqn.~(\ref{simpleMomInertia}) using various EOSs with different
stiffness of symmetry energies.} \label{Fig6}
\end{figure}
\begin{figure} 
\resizebox{0.5\textwidth}{!}{%
\includegraphics{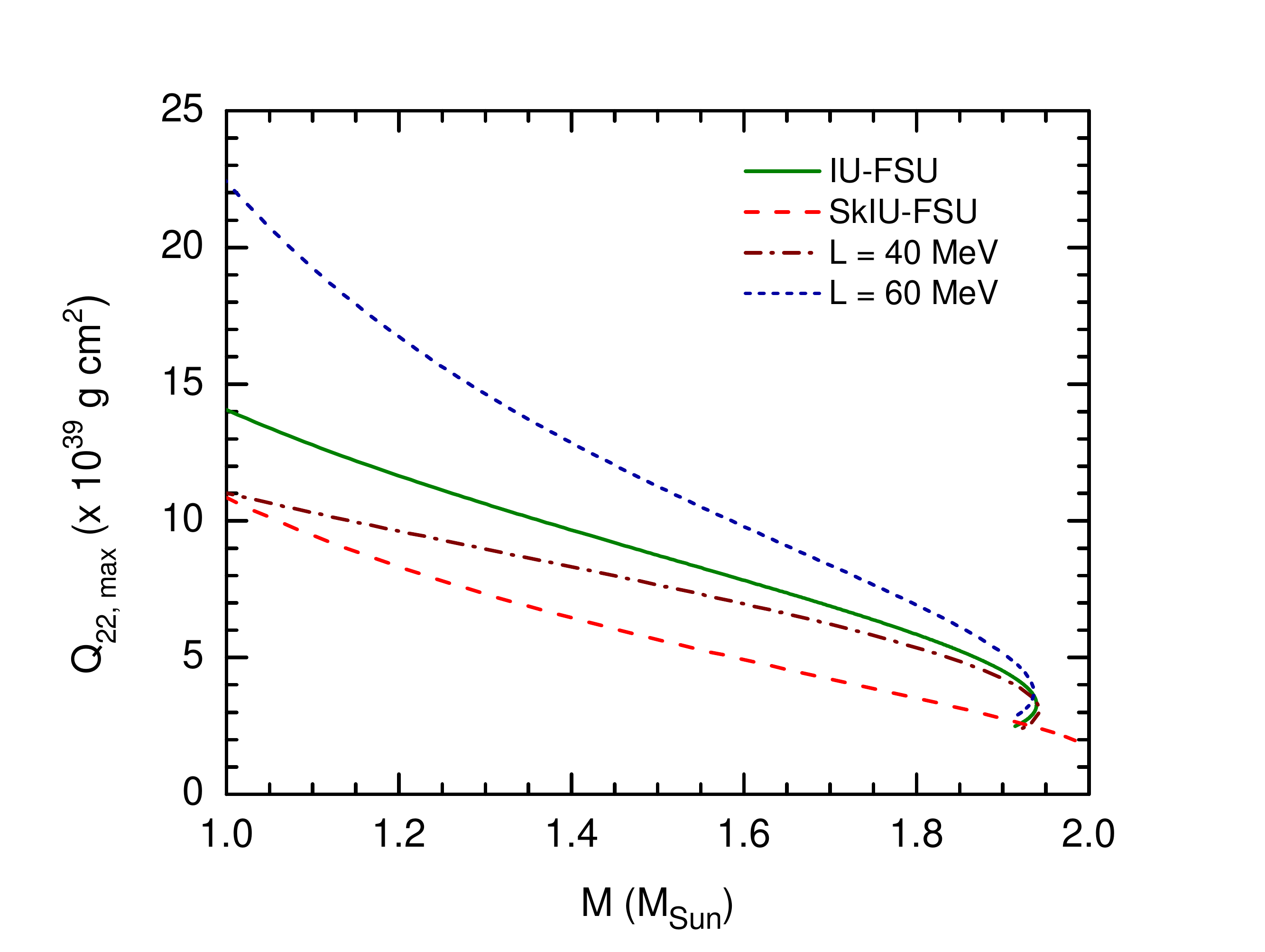}
}
\caption{(Color online) Neutron star maximum quadrupole moment as
calculated using Eqn.~(\ref{quadrupole}).} \label{Fig7}
\end{figure}
In Fig.~\ref{Fig6} we display the neutron star moment of inertia
calculated using a simple empirical relationship proposed by
Ref.~\cite{Lattimer:2004nj}:
\begin{equation}
\label{simpleMomInertia} I \simeq 0.237 MR^2
\left[1+4.2\left(\frac{M \, {\rm km}}{M_{\odot}
R}\right)+90\left(\frac{M \, {\rm km}}{M_{\odot} R}\right)^4\right]
\end{equation}
This expression is shown to hold for a wide class of equations of
state that do not exhibit considerable softening and for neutron
star models above 1 $M_{\odot}$. In Fig.~\ref{Fig7} we display the
maximum quadrupole moment calculated via Eqn.~(\ref{quadrupole}),
where we chose the value of the breaking strain as $\sigma_{\rm max}
= 0.1$. We notice that both moment of inertia and the maximum
quadrupole moment are sensitive to the density dependence of the
symmetry energy. Particularly, low-mass neutron stars exhibit a
strong sensitivity to the density dependence of the nuclear symmetry
energy around saturation density, while massive neutron stars are
more sensitive to the high-density component of the symmetry energy.
When the general relativistic expression for the moment of inertia
is used the result remains the same qualitatively, although it
slightly changes quantitatively. We emphasize that the gravitational
strain amplitude as calculated above is independent of the moment of
inertia, and strongly depends on the axial-asymmetric quadrupole moment.
\begin{figure}[h]
\resizebox{0.5\textwidth}{!}{%
\includegraphics{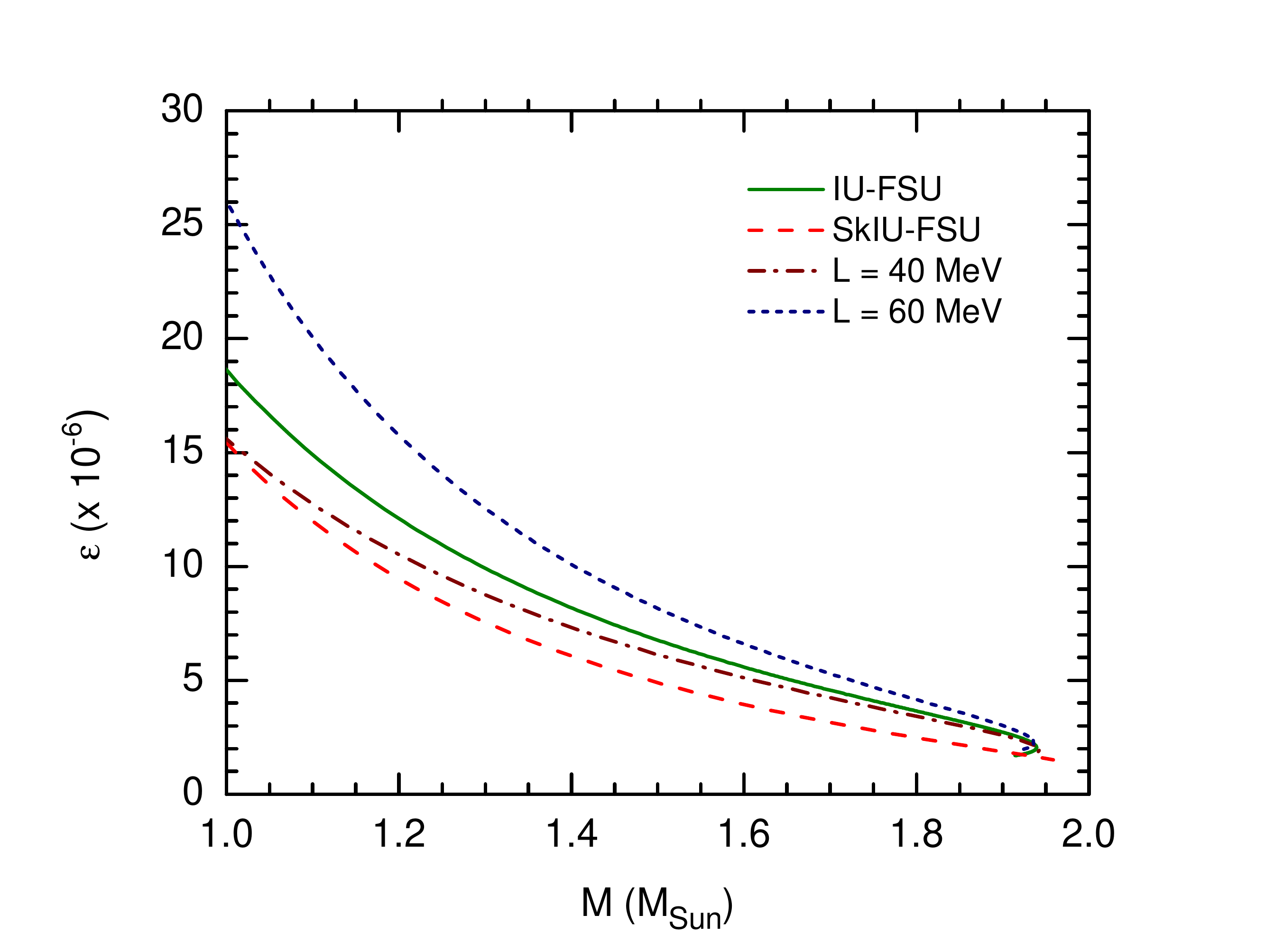}
}
\caption{(Color online) Ellipticity as a function of the neutron
star mass.} \label{Fig8}
\end{figure}
\begin{figure}[h]
\resizebox{0.5\textwidth}{!}{%
\includegraphics{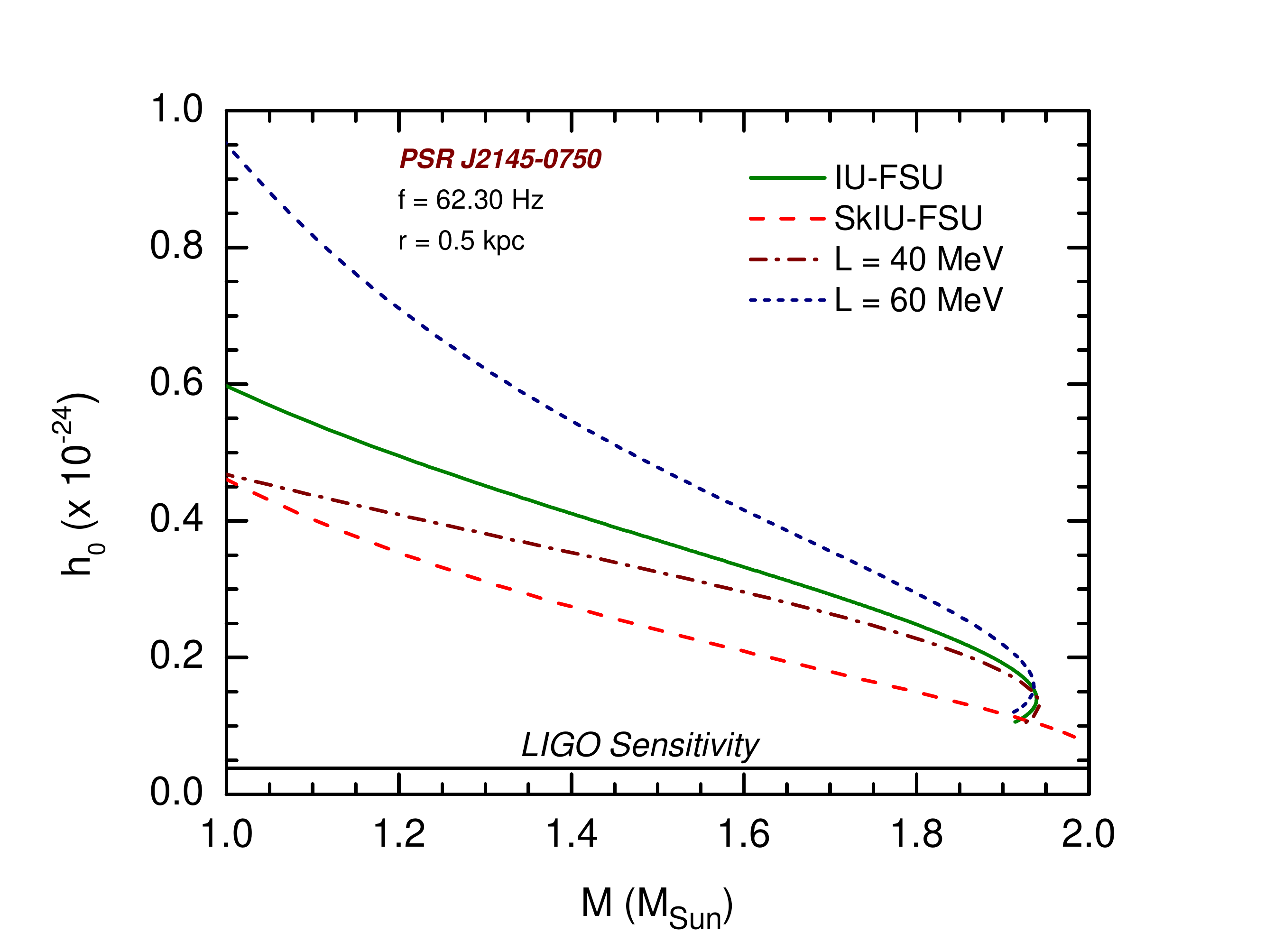}
}
\caption{(Color online) Gravitational wave strain amplitude as a
function of the neutron star mass. The result is shown for pulsar
J2145$-$0750.} \label{Fig9}
\end{figure}
In Fig.~\ref{Fig8} we show the maximum ellipticity that can be supported by the crust as a function of the
stellar mass. Since $\epsilon$ is proportional to the quadrupole
moment scaled by the moment of inertia, it decreases with increasing
stellar mass. It is seen that models with stiff symmetry energy
favor the larger crust ``mountains". There is also a significant
difference for the calculated deformations using models with the
same symmetry energy at saturation, but different high-density
component (IU-FSU and SkIU-FSU) of the symmetry energy.

\begin{table*}
\caption{Properties of the nearby pulsars considered in this study.
The first column identifies the pulsar. The remaining columns are
rotational frequency, distance to Earth, the
observed~\cite{Collaboration:2009rfa} and the calculated upper
limits on the gravitational wave strain amplitude. Notice that the
masses of most of these pulsars are presently unknown and therefore
we have adopted a canonical $1.4 M_{\odot}$ mass neutron star in
calculating the gravitational wave strain amplitudes, which are all
given in units of $1.0 \times 10^{-24}$.}
\label{Tab3}       
\begin{tabular}{lcccccccc}
\hline\noalign{\smallskip}
Pulsar & $f$ $({\rm Hz})$  & $r$ $({\rm kpc})$ & $h_0^{\rm ob}$ & $h_0^{\rm th}$ (IU-FSU) & $h_0^{\rm th}$ (IU-FSU-min) & $h_0^{\rm th}$ (IU-FSU-max) & $h_0^{\rm th}$ (SkIU-FSU)  \\
\noalign{\smallskip}\hline\noalign{\smallskip}
J0437$-$4715 & 173.69 & 0.1 & $0.5730 $ & $15.9590$ & $13.7350$ & $21.2256$ & $10.6567$ \\
J0613$-$0200 & 326.60 & 0.5 & $0.1110 $ & $11.2855$ & $~9.7127$ & $15.0097$ & $~7.5359$ \\
J0751$+$1807 & 287.46 & 0.6 & $0.1640 $ & $~7.2855$ & $~6.2702$ & $~9.6898$ & $~4.8649$ \\
J1012$+$5307 & 190.27 & 0.5 & $0.0694 $ & $~3.8303$ & $~3.2965$ & $~5.0943$ & $~2.5577$ \\
J1022$+$1001 & ~60.78 & 0.4 & $0.0444 $ & $~0.4886$ & $~0.4205$ & $~0.6498$ & $~0.3262$ \\
J1024$-$0719 & 193.72 & 0.5 & $0.0501 $ & $~3.9704$ & $~3.4171$ & $~5.2807$ & $~2.6513$ \\
J1455$-$3330 & 125.20 & 0.7 & $0.0515 $ & $~1.1846$ & $~1.0195$ & $~1.5755$ & $~0.7910$ \\
J1730$-$2304 & 123.11 & 0.5 & $0.0593 $ & $~1.6035$ & $~1.3801$ & $~2.1327$ & $~1.0708$ \\
J1744$-$1134 & 245.43 & 0.5 & $0.1100 $ & $~6.3730$ & $~5.4848$ & $~8.4761$ & $~4.2556$ \\
J1857$+$0943 & 186.49 & 0.9 & $0.0727 $ & $~2.0442$ & $~1.7593$ & $~2.7188$ & $~1.3650$ \\
J2019$+$2425 & 254.16 & 0.9 & $0.0923 $ & $~3.7969$ & $~3.2678$ & $~5.0499$ & $~2.5354$ \\
J2124$-$3358 & 202.79 & 0.2 & $0.0485 $ & $10.8773$ & $~9.3614$ & $14.4668$ & $~7.2634$ \\
J2145$-$0750 & ~62.30 & 0.5 & $0.0383 $ & $~0.4106$ & $~0.3534$ & $~0.5462$ & $~0.2742$ \\
J2322$+$2057 & 207.97 & 0.8 & $0.1120 $ & $~2.8600$ & $~2.4614$ & $~3.8038$ & $~1.9098$ \\
 \noalign{\smallskip}\hline
\end{tabular}
\vspace*{0cm}  
\end{table*}

Finally, in Table~\ref{Tab3} we report calculated upper limits on
gravitational wave strain amplitude for various pulsars and compare
the results with observational upper limits. The results illustrate the relationships discussed above.
It is shown that the gravitational wave strain amplitude is very sensitive to the density
dependence of the symmetry energy. This sensitivity is most apparent
in the Eqn. (\ref{quadrupole}) of the maximum quadrupole moment that
depends on the $6.26$th power of the stellar radius. While
gravitational waves for low-mass neutron stars are sensitive to the
density slope of the nuclear symmetry energy, this sensitivity
weakens as the neutron star becomes more massive. For massive
neutron stars the effect of the high-density component of the
symmetry energy is dominant (Also see Fig.~\ref{Fig9}). The results
in Table~\ref{Tab3} suggest that at present the gravitational
radiation from these pulsars should be within the detection
capabilities of LIGO~\cite{Collaboration:2009rfa}. However, no
signal detection from any of these targets were reported. The fact
that such a detection has not been made yet deserves a few comments.
First and most importantly, recall that in computing the upper
limits on $h_0$ we have used the maximum value for the breaking
strain of the neutron star crust $\sigma =
0.1$~\cite{Horowitz:2009ya}, which might be too optimistic. Even if
the value is right, it is very important to understand that this
does not suggest that neutron stars will have deformations of such
magnitude. In fact, the main problem is to provide a reasonable
scenario that leads to the development of large
deformations~\cite{Andersson:2011}. In this regard,
accretion-powered pulsars could be promising sources due to the
expected asymmetry of the accretion flow near the stellar surface.
However the required modeling for accreting systems have not been
easy because of their complicated dynamics. Besides, none of the
pulsars presented in Table~\ref{Tab3} are known to be in accreting
systems~\cite{PulsarsCat}. Second, we have assumed a fixed $1.4
M_{\odot}$ neutron star, while $h_0$ depends on the neutron star
mass and gets much smaller for massive neutron stars (See Fig.
\ref{Fig9}). Last, distance estimates that are based on dispersion
measure could also be wrong by a factor 2 $-$
3~\cite{Abadie:2010cf}. Obviously, much work remains to be done in
the observational front, in order to extract information on the
density dependence of the nuclear symmetry energy.

\subsection{Neutron star oscillations}
\label{oscillations}

In a similar way that helioseismology studies the internal structure
and dynamics of our Sun, neutron star seismology can be used to
obtain information on various properties of neutron stars, and
therefore to extract the EOS of neutron-rich
matter~\cite{Andersson:1996}. Depending on their characteristics,
neutron-star oscillation modes are usually divided into toroidal and
spherical modes. Further, based on the nature of the restoring force
these modes are classified as follows:
\begin{itemize}
\item[1)] \textit{f}undamental modes, or $f$-modes, associated with the global
oscillation of the fluid, whose frequencies are in the range of 1-10
kHz;
\item[2)] \textit{g}ravity modes, or $g$-modes associated with the fluid buoyancy,
whose frequencies are in the range of 2-100 Hz;
\item[3)]\textit{p}ressure modes, or $p$-modes, associated with pressure gradient and whose frequencies
lie in the range of few kHz;
\item[4)] \textit{r}otational modes (also known as \textit{R}ossby modes), or $r$-modes,
associated with the Coriolis force which acts as a restoring force
along the surface, and whose frequency depends on the stars
rotational frequency;
\item[5)] purely general-relativistic gravitational
\textit{w}ave modes, or $w$-modes, associated with the curvature of
spacetime and typically have a very high frequency of above 7 kHz.
\end{itemize}
There has recently been considerable interest and effort in
extracting information on the density dependence of the symmetry
energy from analyzing various neutron star observations associated
with oscillation modes. In particular, it was shown in
Ref.~\cite{Gearheart:2011,Steiner:2009yg,Sotani:2012,Sotani:2013}
that by identifying quasiperiodic oscillation (QPO) following giant
magnetar flares in soft gamma repeaters with the torsional
oscillations of the crust it may be possible to extract the density
dependence of the symmetry energy (Also see the contributions to
this volume by Iida \& Oyamatsu). For example,
Ref.~\cite{Steiner:2009yg} have calculated the frequency of shear
oscillations of the neutron star crust, and showed that they depend
sensitively on the slope of the symmetry energy at saturation. By
using a consistent treatment of the EOS of crust and core, but an
approximate method for the description of the torsional modes in a
work involved two of us~\cite{Gearheart:2011} we showed that one can
associate the observed QPO frequencies with the fundamental shear
mode, if the density slope of the symmetry energy is $L < 65$ MeV.
Soon after, using a more sophisticated calculation for the crustal
frequencies, but an inconsistent EOS of the crust and core,
Ref.~\cite{Sotani:2012} arrived at a conclusion that the value of
density slope should lie in the range of $100 < L < 130$ MeV. In
their later analysis Ref.~\cite{Sotani:2013} also reported a more
conservative range of $L\geq 47.4$ MeV, depending on the
identification of the fundamental crust modes. These results suggest
that such oscillations could be a powerful probe of the symmetry
energy at saturation density, but that it is important to
incorporate a sophisticated model of the crust, consistent
calculations of the EOS of crust and core and the effects of the
high magnetic fields into the analysis.

Due to their purely general-relativistic nature, $w$-modes have also
become the focal point of many
investigations~\cite{Wen:2009iz,Lin:2010zw}. In particular, it was
shown that the density dependence of the nuclear symmetry energy has
a clear imprint on both the frequency and the damping time of the
axial $w$-modes~\cite{Wen:2009iz}. Although the $w$-mode frequencies
are outside the bandwidth of the current gravitational wave
detectors, major upgrades will be completed over the coming years
that will significantly improve sensitivity required to detect
gravitational waves over a much broader
band~\cite{AdvLigo,EinsteinTelescope}.

In this contribution
we will concentrate on the $r$-mode of the oscillation, whose
study has increased significant attention soon after their first
relativistic calculations~\cite{Andersson:1998,Friedman:1998}.


\subsubsection{The $r$-mode instability in neutron stars}

Theoretically, rotating neutron stars cannot have a spin frequency
larger then their Kepler frequency, $\Omega_{\rm K}$. This
is an absolute upper limit on the stars rotation rate above which
the matter gets ejected from the star's equator. However, the available
observational data suggests that the spin-up of neutron stars from accretion
is limited to frequencies much lower $\Omega/\Omega_{\rm K} <
0.1$. One possible explanation is
the Chandrasekhar-Friedmann-Schutz (CFS) instability of the $r$-mode
oscillations. This instability might play an important role in generating
gravitational waves from the accretion-powered millisecond pulsars
in low-mass X-ray binaries (LMXBs). The CFS instability sets an upper
limit on the rotation frequency, above which $r$-modes are unstable
to gravitational radiation. The $r$-mode instability window
depends on the competition between the gravitational radiation and
the viscous dissipation timescales, where gravitational radiation
makes the $r$-mode amplitude grow and viscosity in the fluid damps the amplitude, stabilizing the mode. It was
first shown in Refs.~\cite{Lindblom:1998,Andersson:1998} that for
all slowly rotating neutron stars, gravitational radiation
timescales exceed the one due to viscous damping. In what follows we
will use the same approach as discussed in
Ref.~\cite{Wen:2011xz,Bildsten:2000,Andersson:2000,Lindblom:2000}.
By assuming that the crust of the neutrons star is perfectly rigid,
one can find an upper limit on the instability window. This is because
the viscous boundary layer between the fluid core and the rigid
crust is maximally dissipative. In the realistic case of more
elastic crust, the dissipation from the core-crust boundary
decreases and therefore widens the instability window. In a recent
work involving some of us~\cite{Wen:2011xz}, we studied the
sensitivity of the $r$-mode instability to the density dependence of
the symmetry energy. In particular, by employing a simple model of a
neutron star with a perfectly rigid crust constructed with a set of
crust and core EOS that span the range of nuclear experimental
uncertainly in the symmetry energy, we found that EOSs characterized
by the density slope $L$ of smaller than $65$ MeV are more
consistent with the observed frequencies in LMXBs. In this work the
electron-electron scattering was considered as the main dissipative
mechanism. However, considering a different approach in which the viscous dissipation
was assumed to operate throughout the whole core of the star instead of at the crust-core boundary layer, and using both microscopic and
phenomenological approaches to the nuclear EOS,
Ref.~\cite{Vidana:2012ex} concluded that observational data seem to
favor values of density slope larger than $50$ MeV. We should make the caveat that both studies use greatly simplified models of the instability, and rely on an interpretation of observations
that is only one of several possible.

According to Ref.~\cite{Lindblom:1998} the gravitational radiation
timescale can be evaluated using the following expression:
\begin{eqnarray}
\label{rmodeGW} \nonumber \frac{1}{\tau_{\rm GR}} &=& \frac{32 \pi G
\Omega^{2l+2}}{c^{2l+5}} \frac{(l-1)^{2l}}{[(2l+1)!!]^2}
\left(\frac{l+2}{l+1}\right)^{2l+2} \ \\ &\times& \int_0^{R_{\rm t}}
\mathcal{E} r^{2l+2} dr \ ,
\end{eqnarray}
where $\Omega$ is the stellar angular frequency and $\mathcal{E}
\equiv \mathcal{E}(r)$ is the energy density profile of the star.
The viscous damping timescale due to viscous dissipation at the
core-crust boundary layer assuming a perfectly rigid crust and fluid
core can be calculated using~\cite{Lindblom:2000}
\begin{equation}
\label{rmodeV} \tau_{\rm v} = \frac{2^{l+1} (l+1)!}{l(2l+1)!!
\mathcal{I}_l c R_{\rm t}^{2l+2}\sqrt{ \mathcal{E}_{\rm t}\Omega
\eta_{\rm t}}}   \int_0^{R_{\rm t}} \mathcal{E} r^{2l+2} dr \ .
\end{equation}
where the integral
\begin{equation}
\mathcal{I}_l = \int_0^{\pi}\sin^{2l-1} \theta (1+\cos \theta)^2
\sqrt{|\cos \theta - 1/(l+1)|} d \theta \ ,
\end{equation}
and the subscript ``$\rm t$" is used to identify quantities at the
core-crust interface. Here we only consider quadrupole modes of
$l=2$, for which $\mathcal{I}_2 \approx 0.80411$. For hot neutron
stars with temperature above $T = 10^9$ K, it is expected that the
neutron-neutron scattering becomes the dominant dissipation
mechanism, and its viscosity is expressed by
\begin{equation}
\eta_{\rm nn} = k_{\rm nn} \rho^{9/4} T^{-2} \ ,
\end{equation}
where
\begin{equation}
k_{\rm nn} = 347 \quad {\rm g}^{-\frac{5}{4}} \, {\rm
cm}^{\frac{23}{4}} \, {\rm K}^{2} \, {\rm s}^{-1} \ ,
\end{equation}
and $\rho$ is the mass density. As the temperature drops below about
$10^9$ K, the neutron star shear viscosity is then dominated by the
electron-electron scattering, which also depends on density and
temperature:
\begin{equation}
\eta_{\rm ee} = k_{\rm ee} \rho^{2} T^{-2} \ ,
\end{equation}
where
\begin{equation}
k_{\rm ee} = 6.0 \times 10^6 \quad {\rm g}^{-1} \, {\rm cm}^{5} \,
{\rm K}^{2} \, {\rm s}^{-1} \ ,
\end{equation}
The critical rotation frequency $\Omega_{\rm c}$ is then defined
when $\tau_{\rm GR} = \tau_{\rm v}$. In this work we do not consider
new-born stars, whose temperatures are usually above $T > 10^{9}$ K
and for which the bulk viscosity would be the dominant dissipation
mechanism.
\begin{figure}[h]
\resizebox{0.5\textwidth}{!}{%
\includegraphics{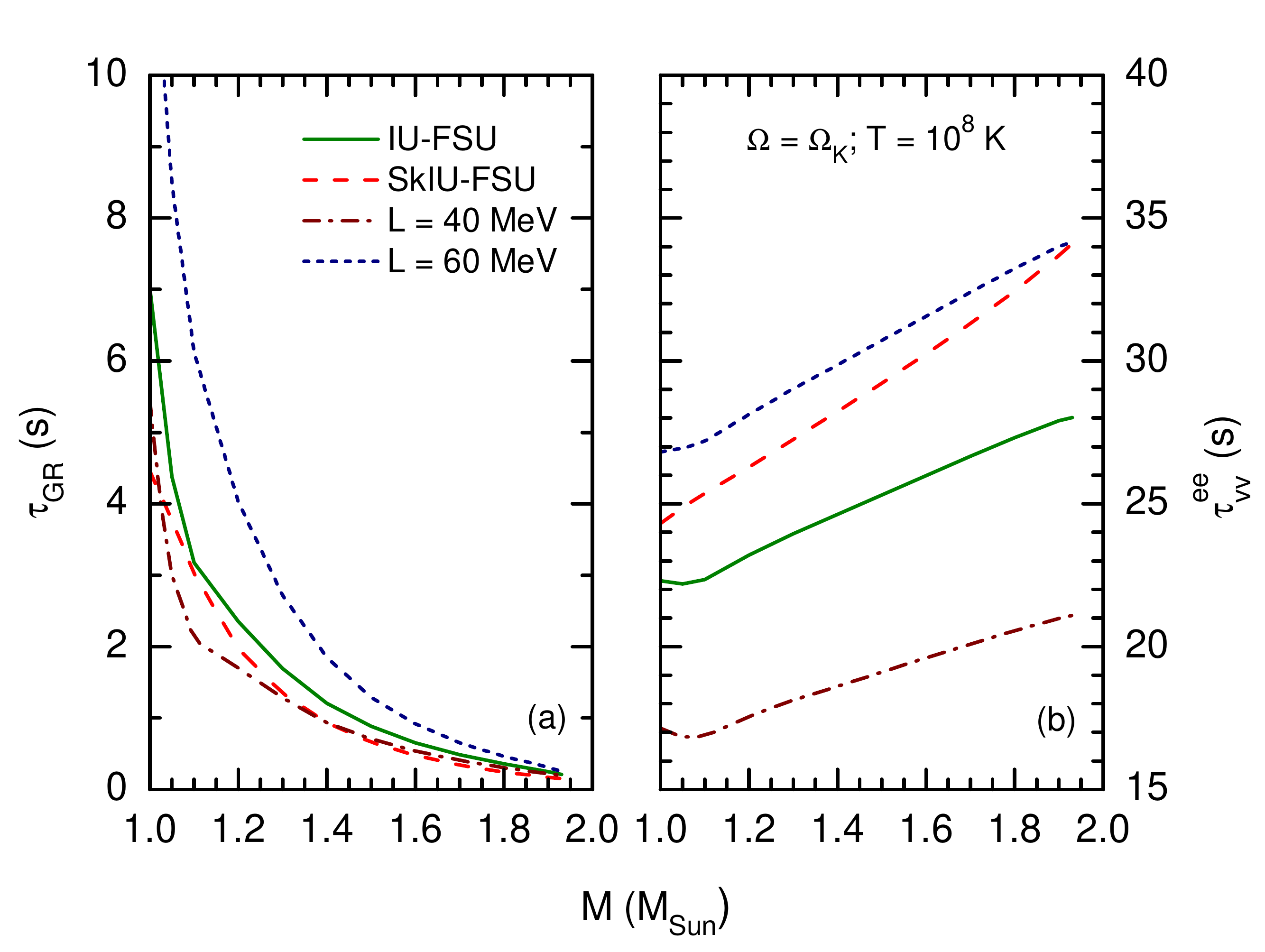}
}
\caption{(Color online) Timescales for the (a) gravitational
radiation driven $r$-mode instability and (b) corresponding viscous
dissipation due to the electron-electron scattering at the
core-crust interface as a function of stellar mass.} \label{Fig10}
\end{figure}

In Fig.~\ref{Fig10} the timescales for the gravitational radiation
(left window) and the shear viscosity (right window), which
dissipates the $r$-mode instability are shown as a function of
stellar mass. The stellar structure is modeled with a rigid
crust~\cite{Wen:2011xz} for the adopted EOSs, and the shear
viscosity of the boundary layer is taken to be dominated by
electron-electron scattering. Both timescales depend on the stellar
mass with gravitational timescale being a decreasing function of the
neutron star mass, whereas the viscous damping timescale is an
increasing function of the stellar mass. Although both the
gravitational radiation timescale and the viscous damping timescale
seems to be more sensitive to the density dependence of the symmetry
energy at saturation, an apparent sensitivity emerges for the
high-density component of the symmetry energy in the viscous damping
timescale only. Indeed, we should note that these timescales
strongly depend on the core-to-crust transition properties, which
occurs at about half saturation density. The sensitivity to the high
density component of the symmetry energy only enters through the
integrals shown in Eqns. (\ref{rmodeGW})-(\ref{rmodeV}) which are evaluated numerically.
\begin{figure}[h]
\resizebox{0.5\textwidth}{!}{%
\includegraphics{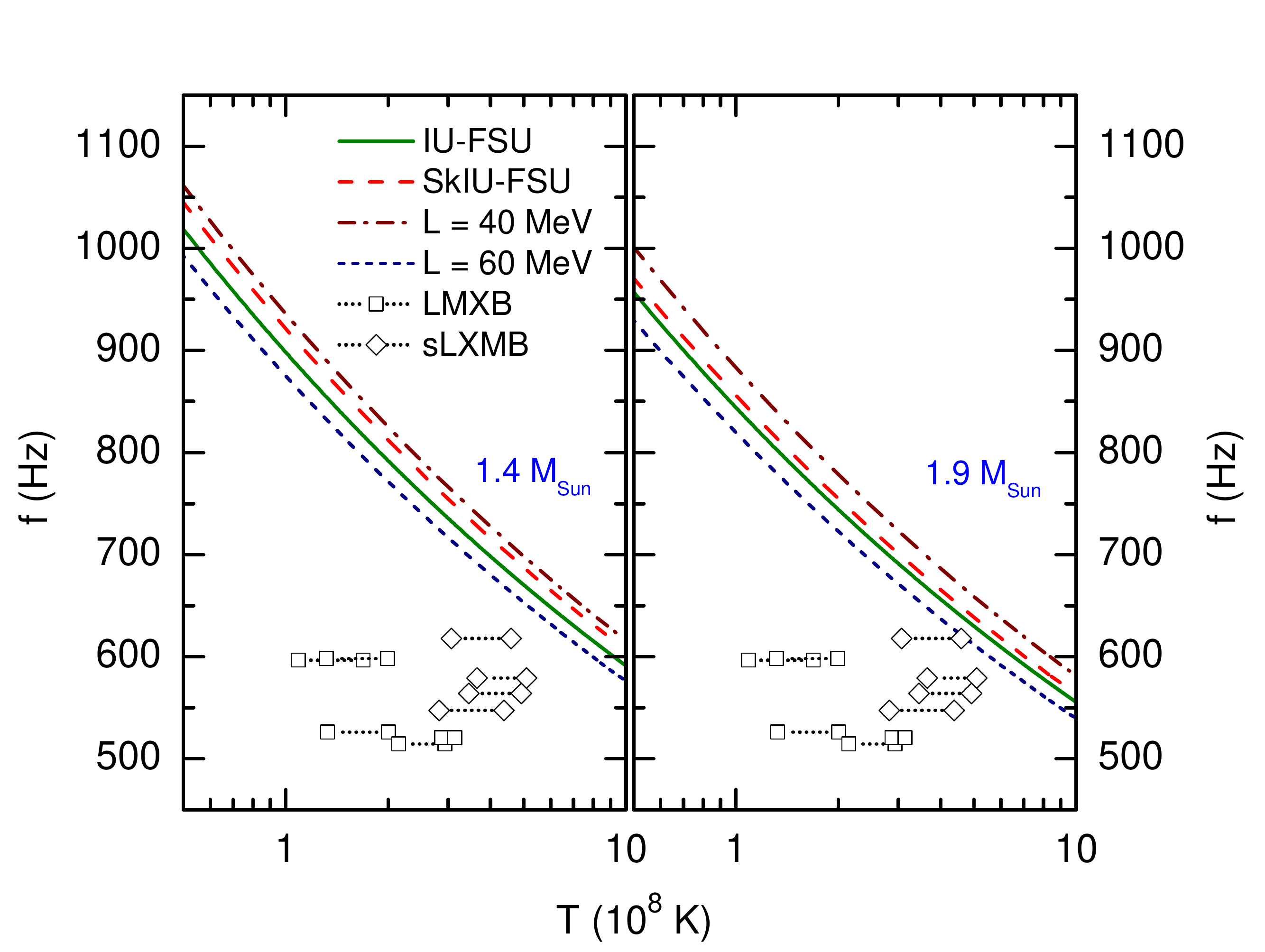}
}
\caption{(Color online) The $r$-mode instability window for a
$1.4$-solar mass (left window) and a $1.9$-solar mass neutron stars
are shown for various EOSs discussed in the text. Also the location
of the observed low mass X-ray binaries (LMXBs) and short recurrence
time LMXBs (sLMXB) are shown~\cite{Ho:2011}. } \label{Fig11}
\end{figure}

Since the masses of the neutron stars in low-mass X-ray binaries
(LMXBs) are not measured as accurately as those in certain binary
NS-NS or NS-White Dwarf (WD) systems, we explore the position of
LMXBs in the $r$-mode instability window, defined as the region in
frequency-temperature space above the critical frequency, in both
canonical and massive neutron stars. The $r$-mode instability window
for neutron stars of $1.4 M_{\rm Sun}$ and $1.9 M_{\rm Sun}$ are
displayed in the left and right windows of Fig.~\ref{Fig11},
respectively. The core temperatures $T$ of the LMXBs are derived
from their observed accretion luminosity assuming the cooling is
either dominated by the modified Urca process for normal nucleons
(left stars) or by the modified Urca process for neutrons being
superfluid and protons being superconducting (right stars) in the
core~\cite{Ho:2011}. It is shown that for a canonical neutron star,
all considered LMXBs lie outside the instability window and is
consistent with the observation that no $r$-mode is currently
excited in LMXBs, due to the shortness in duration of the unstable
$r$-mode activity~\cite{Levin:1999,Bondarescu:2007}. However, for
the massive neutron stars, some of the stars fall within the
instability region constrained by the EOSs. Considering that LMXBs
should fall below the instability window and assuming that current
observational interpretation is correct, then within our simple
model one can conclude from Fig.~\ref{Fig11} that one of the
following must hold: (a) stars in LMXBs are not so massive; and
either (b) the high-density component of the symmetry energy is
soft, or (c) the density slope of the symmetry energy at saturation
is $L < 60$ MeV. Suppose that the short recurrence time LMXBs
(sLMXBs) presented in the Fig.~\ref{Fig11} are identified to be
massive, and the density slope of the symmetry energy $L$
constrained by terrestrial experiments is equal or larger then $60$
MeV, then one would conclude that the symmetry energy must be soft
at higher densities.


\section{Conclusions}
\label{conclusions}

In this survey we have examined the sensitivity of various neutron
star-sourced gravitational wave signals to the density dependence of
symmetry energy, with a special emphasis on the high density
behavior of the symmetry energy. Specifically, we have addressed the
sensitivity of gravitational wave signals from tidally polarized
neutron stars, accretion-induced neutron star ellipticities, and
$r$-mode oscillations to the EOSs of neutron-rich matter.

To study this sensitivity we have used various EOSs for neutron-rich
nucleonic matter satisfying the latest constraints from terrestrial
nuclear experiments, state of the art nuclear many-body calculations
for EOS of PNM, and astrophysical observation. We have used this
same set of EOSs to test each of the different GW sources, so that
their relative sensitivities to dense nuclear matter properties may
be consistently compared.

We have found that among various gravitational wave signatures the
tidal polarizability of neutron stars in coalescing binaries is
particularly sensitive to the high-density behavior of the nuclear
symmetry energy. Moreover, tidal polarizability is relatively
insensitive to variations of the EOS of SNM and the density
dependence of the symmetry energy around saturation density within
their remaining experimental and model uncertainty ranges.

Next, we calculated the gravitational wave strain amplitude from
neutron stars elliptically deformed by crustal mountains. The exact
calculation of the neutron star quadrupole moment (and therefore the
gravitational wave strain amplitude) requires a detailed knowledge
of the neutron star crust. Within a simple approximation, our
estimations show that it may be possible to disentangle the effect
of the high-density component of the symmetry energy from its
density dependence around saturation density for the very massive
neutron stars only. Much works need to be done to completely
understand the physics of neutron star crust, in particular its
ability to support large ellipticities, but it is clear that
observations of gravitational waves from accreting neutron stars is
a promising avenue towards constraining the EOS of neutron-rich
nucleonic matter.

Lastly, we have explored the dependence of the $r$-mode instability
window on the high-density component of the symmetry energy. We have
analyzed and confirmed the previous findings that the instability
window is mostly sensitive to the EOS at around the crust-core
transition density, and therefore to the density dependence of the
symmetry energy around saturation density. The knowledge of high
density behavior of the symmetry energy becomes very important,
especially when one combines future stringent experimental
constraints on the value of $L$ with existing and future
observations of LMXBs. Within this simple model, we have found that
EOSs characterized by $L$ value smaller then 60 MeV are consistent
with the observations. In particular, if large values of $L$ are
favored from forthcoming terrestrial experiments, then models with
stiff symmetry energy at high densities would be ruled out.

Future measurements of the gravitational wave signals will therefore
be utterly important in constraining stringently the high-density
behavior of nuclear symmetry energy, and thus the nature of dense
neutron-rich nucleonic matter.

\section*{Acknowledgments}
\label{acknowledgments} This work was supported in part by the
National Aeronautics and Space Administration under Grant No.
NNX11AC41G issued through the Science Mission Directorate, and the
National Science Foundation under Grant No. PHY-1068022.
%
%
%

\end{document}